\documentclass[11pt,twoside]{article}

\usepackage{asp2006}
\usepackage{epsf}
\usepackage{psfig}
\usepackage{lscape}

\usepackage{graphicx}

\begin{document}
\title{Mid-infrared observations of NGC~1068 with VLT/VISIR}   
\author{Poncelet, A.\altaffilmark{1,2}, Doucet, C.\altaffilmark{3},
  Perrin, G.\altaffilmark{2}, Sol, H.\altaffilmark{1} and Lagage, P.O.\altaffilmark{3}}

\altaffiltext{1}{LUTH, Paris Observatory, 5 place Jules Janssen, 92195
Meudon Cedex}
\altaffiltext{2}{LESIA, Paris Observatory, 5 place Jules Janssen, 92195
Meudon Cedex}
\altaffiltext{3}{CEA/DSM/DAPNIA/Service d'Astrophysique, CE Saclay
  F-91191 Gif-sur-Yvette}

\begin{abstract} 
We present a speckle analysis of the active galactic nucleus (AGN) inside the archetype Seyfert type 2 galaxy
NGC~1068. This study is based on 12.8~$\mu$m images obtained with the burst
mode of VISIR (the Very Large Telescope Imager and Spectrometer
in the InfraRed). The interferometric processing allows to push the resolution
far below the diffraction limit of a 8~m telescope in the
$N$~band and to trace two main contributions to the mid-IR flux inside
the nucleus. It also allows to partially fill the lack of visibility points at low
spatial frequencies. The confrontation with VLT/MIDI (the
Mid-InfrareD Interferometer) data points helps to establish the link between dust in the
vicinity of the central engine and inside the ionisation cone to
get a multi-scale picture of mid-IR sources emitting in the
nucleus of NGC~1068.  
\end{abstract}





\section{Speckle analysis of VISIR BURST mode images \label{principle}}

The speckle processing of 12.8~$\mu$m VISIR burst mode images aims
to fully benefit of the diffraction limit of a UT (Unit Telescope of
8~m at the VLT). Low spatial frequency visibilities between 0 and 8~m
(refered as \textit{VISIR visibilities}) are obtained from the Fourier
transform of the source distribution of intensity.

\begin{figure}[ht!]
 \centering \includegraphics[height=4.5cm]{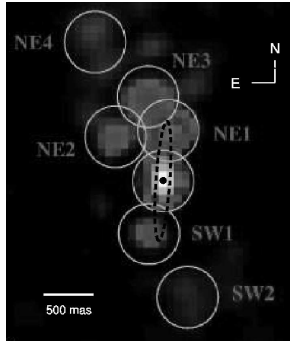}
 \caption{Superimposition of structures traced by the
  speckle processing (black inner rim and dotted elongated ellipse) on the
  deconvolved 12.8~$\mu$m VISIR image of \citet{Galliano}. Our ellipse must be understood as an average model of the knots traced
  by deconvolution and this picture shows the good consistency of both techniques.}
 \label{fig1} 
\end{figure}

The application of a model (containing \textit{a priori}
information on the source) allows to extrapolate high-spatial
frequency visibilities from the low spatial frequency ones and then to estimate structures smaller
than the diffraction limit ($\sim$~320~mas at 12.8~$\mu$m). 
In this way, the model accounting for the \textit{VISIR visibilities} considers two main contributions to the MIR flux: (i) a small
component of $<$~85~mas, well below the diffraction limit and in
full agreement with the sizes of the dusty torus from \citet{Jaffe} and
\citet{Poncelet_2006a}; (ii) a NS elongated one at P.A.~=~-4$^\circ$,
$<$~140~mas~$\times$~1197~mas \citep{Poncelet_2006b}, is associated to knots NE1 and
SW1 from deconvolved images of VISIR standard mode images
\citep{Galliano} (see Fig.~\ref{fig1}). It shows the good consistency between the two
approaches.

\section{The link between VISIR and MIDI visibilities \label{link}}

\begin{figure}[ht!]
 \centering \includegraphics[height=4cm]{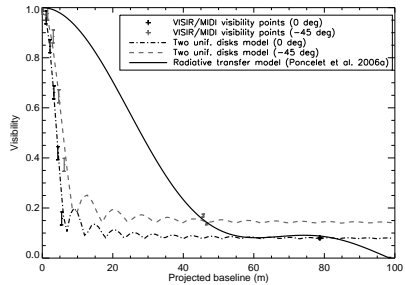}
 \caption{Confrontation between [0-8]~m visibilities from the speckle analysis
  of VISIR images and long baselines MIDI data points at
  12.8~$\mu$m. They are compared to the radiative transfer model of \citet{Poncelet_2006a} (black line). A model of two components of $\sim$~20~mas and
  $<$~400~mas is mandatory to reproduce both data sets together \citep{Poncelet_2006b}. }
 \label{fig2} 
\end{figure}

To confront [0~-~8]~m \textit{VISIR visibilities} with MIDI data of 2003,
we take into account the limited field of view of MIDI
(i.e. a 0.6$''~\times$~0.6$''$ area oriented at -30$^\circ$) in the speckle processing.
Fig.~\ref{fig2} presents the confrontation between low spatial frequency
visibilities then obtained and MIDI data points of 2003. It illustrates
the strong and unexpected drop of visibilities at short
baselines. Both data sets are well reproduced by a model of two
components entering in the field of view of MIDI: a compact one
($\sim$~20~mas) associated to the \textit{dusty layer}, and an
extended one ($<$~400~mas) associated with heated dust in the
ionisation cone \citep{Poncelet_2006b}. 
This amount of dust of the ionisation cone entering in the field of
view of MIDI has then to be taken into account during the modelling of
interferometric data.

\end{document}